\def\SGMPcomment#1{}

\def\SGMPdefifnonempty#1#2{\def\tempcs{#2}%
	\ifx \tempcs\empty \else
		\def#1{#2}%
	\fi
}

\let\mdef=\SGMPdefifnonempty

\newif\ifSGMPkeeplooping

\newtoks\SGMPttlpgtoks	\SGMPttlpgtoks={}

\long\def\SGMPttlpgmisc#1{\relax
	\expandafter\SGMPttlpgtoks\expandafter{\the\SGMPttlpgtoks
				\vskip6pt plus6pt minus4pt#1\par}%
}

\def\SGMPpreface{%
	\edef\pgn{\the\count0}\begin{titlepage}
	\pagestyle{empty} \vglue1.0in\leftline{\huge Preface}
	\vskip4pc\normalsize\noindent
}

\def\SGMPendpreface{%
	\end{titlepage}\global\advance\count0 by\pgn
	\global\advance\count0 by 1
}

\catcode`\@=11
\def\SGMPmaketitle{%
	\gdef\@thanks{\begin{center}\normalsize\the\SGMPttlpgtoks\end{center}}%
	\maketitle
}

\def\SGMPref#1{%
	\@ifundefined{r@x:#1}{\pageref{pg:#1}}{\ref{x:#1}}%
}
\catcode`\@=12

\def\SGMPunident{}
\def\SGMPlab#1{\ifx#1\SGMPunident\else\label{x:#1}\fi}

\newtoks\SGMPstackA		\global\SGMPstackA={}
\newtoks\SGMPstackB		\global\SGMPstackB={}

\def\push #1\onto#2{\relax
  \global\SGMPstackA=\expandafter{#1}\global\SGMPstackB=\expandafter{#2}%
  \xdef#2{{\the\SGMPstackA}{\the\SGMPstackB}}%
}

\def\gsetSGMPstackB{\global\SGMPstackB}

\def\pop #1\from#2{\relax
  \afterassignment\gsetSGMPstackB \global\SGMPstackA=#2{}{}%
  \xdef#1{\the\SGMPstackA}\xdef#2{\the\SGMPstackB}%
}

\def\SGMPbeginList#1#2{%
	\def\SGMPlisttype{#1}%
	\push \SGMPendList\onto \SGMPendListstack
	\push \SGMPitem\onto \SGMPitemstack
	\ifx \SGMPlisttype\SGMPlisttext
		\def\SGMPendList{%
			\end{description}%
			\pop \SGMPendList\from \SGMPendListstack
			\pop \SGMPitem\from \SGMPitemstack
		}%
		\def\next{\begin{description}}%
		\def\SGMPitem{\item[#2]}%
	\else \ifx \SGMPlisttype\SGMPlistnone
		\def\SGMPendList{%
			\end{description}%
			\pop \SGMPendList\from \SGMPendListstack
			\pop \SGMPitem\from \SGMPitemstack
		}%
		\def\next{\begin{description}}%
		\def\SGMPitem{\item[]}%
	\else \ifx \SGMPlisttype\SGMPlistbulleted
		\def\SGMPendList{%
			\end{itemize}%
			\pop \SGMPendList\from \SGMPendListstack
			\pop \SGMPitem\from \SGMPitemstack
		}%
		\def\next{\begin{itemize}}%
		\def\SGMPitem{\item}%
	\else \ifx \SGMPlisttype\SGMPlistsquare
		\def\SGMPendList{%
			\end{itemize}%
			\pop \SGMPendList\from \SGMPendListstack
			\pop \SGMPitem\from \SGMPitemstack
		}%
		\def\next{\begin{itemize}}%
		\def\SGMPitem{\item[$\Box$]}%
	\else
		\def\SGMPendList{%
			\end{enumerate}%
			\pop \SGMPendList\from \SGMPendListstack
			\pop \SGMPitem\from \SGMPitemstack
		}%
		\def\next{\begin{enumerate}}%
		\def\SGMPitem{\item}%
	\fi \fi \fi \fi
	\next
}

\let\SGMPendList=\relax		\let\SGMPitem=\relax
\let\SGMPendListstack=\relax	\let\SGMPitemstack=\relax
\def\SGMPlisttext{text} \def\SGMPlistnone{none}
\def\SGMPlistbulleted{bulleted} \def\SGMPlistsquare{square}

\def\SGMPcite#1#2{\cite{#1}}

\makeindex
\def\SGMPindex#1#2{\def\tempcs{#1}%
	\ifx \tempcs\SGMPvalueyes
		#2%
	\fi
	\index{#2}%
}
\def\SGMPvalueyes{yes}
\begingroup \catcode`\@=11
\newread\dfe@ \gdef\dfe#1#2#3{\relax
       \immediate\openin\dfe@=#1 \ifeof\dfe@#3\else#2\fi
       \immediate\closein\dfe@}
\gdef\SGMPstartindex{\relax\ifx\@indexfile\undefined\else
       \closeout\@indexfile \fi\begin{theindex}
       \def\indexentry##1##2{\item##1 ##2}}
\gdef\SGMPfinishindex{\dfe{\jobname.ind}{\def\next{\input
       \jobname.ind}}{\let\next=\relax}\ifx\next\relax \dfe
       {\jobname.idx}{\def\next{\input \jobname.idx}}{\relax}\fi
       \next \end{theindex}}
\endgroup

\def\aalign#1{\leavevmode\vbox{\baselineskip=0pt \lineskiplimit.25ex
  \ialign{##\crcr#1\crcr}}}
\def\SGMPring#1{\aalign{\hidewidth\char"17\hidewidth\cr\noalign{\kern-1.2ex}#1}}

\let\SGMPnewline=\\

\newtoks\TexMacPairEndtextoks

\def\SGMPgobble#1{}

\def\SGMPlim#1{\def\tempcs{#1}%
	\ifx \tempcs\empty
		\let\SGMPdolim=\displaylimits
	\else \if #1c
		\let\SGMPdolim=\limits
	\else \if #1r
		\let\SGMPdolim=\nolimits
	\else
		\let\SGMPdolim=\relax
	\fi \fi \fi
}

\def\Rad#1{%
	\begingroup
	\def\RadTempCs{{#1}}\let\RdxTempCs=\empty
}

\def\DoRad{%
	\relax
	\ifx \RdxTempCs\empty
		\sqrt\RadTempCs
	\else
		\root \RdxTempCs \of \RadTempCs
	\fi
	\endgroup
}

\def\LeftPost#1{\csname LP#1\endcsname}
\def\RightPost#1{\csname RP#1\endcsname}

\def\getchar #1#2\endgetchar{\def\gotchar{#1}\def\ungotchars{#2}}

\def\SGMPmathgrk#1{%
    \def\ungotchars{#1}%
    \SGMPkeeploopingtrue
    \loop
	\expandafter\getchar\ungotchars\endgetchar
	\ifx \gotchar\empty \def\gotchar{0}\fi
	\count255=\expandafter`\gotchar\relax
	\advance\count255 by -49
	\ifcase \count255
		\nabla
	\or	\varpi
	\or	\varepsilon
	\or	\varphi
	\or		
	\or	\partial
	\or\or
	\or	\varrho
	\or\or\or\or\or\or\or
	\or	A%
	\or	B%
	\or	X%
	\or	\Delta
	\or	E%
	\or	\Phi
	\or	\Gamma
	\or	H%
	\or	I%
	\or		
	\or	K%
	\or	\Lambda
	\or	M%
	\or	N%
	\or	O%
	\or	\Pi
	\or	\Theta
	\or	P
	\or	\Sigma
	\or	T%
	\or	\Upsilon
	\or
	\or	\Omega
	\or	\Xi
	\or	\Psi
	\or	Z%
	\or\or\or\or\or\or
	\or	\alpha
	\or	\beta
	\or	\chi
	\or	\delta
	\or	\epsilon
	\or	\phi
	\or	\gamma
	\or	\eta
	\or	\iota
	\or	\vartheta
	\or	\kappa
	\or	\lambda
	\or	\mu
	\or	\nu
	\or	o%
	\or	\pi
	\or	\theta
	\or	\rho
	\or	\sigma
	\or	\tau
	\or	\upsilon
	\or	\varsigma
	\or	\omega
	\or	\xi
	\or	\psi
	\or	\zeta
	\else
	\fi
	\relax
	\ifx \ungotchars\empty \SGMPkeeploopingfalse \fi
	\ifSGMPkeeplooping
    \repeat
}

\catcode`\@=11
\def\eqalign#1{\null\,\vcenter{\openup\jot\m@th
  \ialign{\strut\hfil$\displaystyle{##}$&$\displaystyle{{}##}$\hfil
      \crcr#1\crcr}}\,}
\catcode`\@=12

\def\MthAcnt#1#2{#2{#1}}

\def\SGMPgraphic#1#2#3#4#5#6#7{{%
	\def\type{#3}
	\def\imresdefault{#4}
	\def\imresvalue{#5}
	\def\picresdefault{#6}
	\def\picresvalue{#7}

	\def\yes{yes}
	\def\drawing{drawing}
	\def\image{image}
	\def\height{4in}

	\ifx\type\drawing	
	    \vbox to\height{%
			\special{pub: pubdraw #2 #10}
			\vfil}
	\else\ifx\type\image	
	    \ifx\imresdefault\yes
		\vbox to\height{%
			\vfil
			\special{pub: sunbitmap #2 #10 0}}
	    \else
		\vbox to\height{%
			\vfil
			\special{pub: sunbitmap #2 #10 \imresvalue}}
	    \fi
	\else			
	    \ifx\picresdefault\yes
		\vbox to\height{%
			\vfil
			\special{pub: sunbitmap #2 #10 0}}
	    \else
		\vbox to\height{%
			\vfil
			\special{pub: sunbitmap #2 #10 \picresvalue}}
	    \fi
	\fi\fi
}}

\ifx\TMPcountA\undefined
  \catcode`\!=10
\else
  \catcode`\!=14
\fi
!\newcount\TMPcountA
!\newcount\TMPcountB
!\newdimen\TMPdimenA
!\newdimen\TMPdimenB
\catcode`\!=12

\def\postscript#1#2#3#4#5#6{%
  \TMPdimenA=#5\relax
  \TMPdimenB=#6\relax
  \TMPcountA=\TMPdimenA
  \TMPcountB=\TMPdimenB
  \hbox to #1{%
    \vbox to #2{
      \vss
      \special{ps: plotfile #3 asis}
      \special{ps::[asis,end]
         ChartCheckPoint restore
         0 SPE
      }
    }%
    \hss
  }%
}

\def\SGMPTabcnvtlist#1#2{%
	\def\tempcsA{#1}%
	\def#2{}%
	\ifx \tempcsA\empty \else
	    \SGMPkeeploopingtrue
	    \loop
		\expandafter\SGMPparsetablelist\tempcsA:::\endSGMPparsetablelist#2
		\ifx \tempcsA\empty
			\SGMPkeeploopingfalse
		\fi
		\ifSGMPkeeplooping
	    \repeat
	\fi
}
\def\SGMPparsetablelist #1:#2::#3\endSGMPparsetablelist#4{%
	\def\tempcsA{#2}%
	\expandafter\def\expandafter#4\expandafter{#4\\#1}%
}

\def\SGMPTabColW#1{\SGMPTabcnvtlist{#1}\TabColW}

\def\SGMPTableWd#1{\def\tempcs{#1}%
	\ifx \tempcs\SGMPabs
		\def\TableWd{A}%
	\else
		\def\TableWd{R}%
		\def\TableWdRPct{#1}%
	\fi
}

\def\SGMPabs{abs}

\def\SGMPbeginTable#1#2#3#4#5#6#7#8#9{%
	\edef\TabRuleVO{\TabRuleVI}\edef\TabRuleHO{\TabRuleHI}%
	\SingleRuleWidthInPixels=6
	\Table[\SGMPTableWd{#9}\mdef\TableJust{#6}\SGMPTabColW{#5}%
		\mdef\TabJustVO{#8}\mdef\TabJustVH{#8}%
		\mdef\TabRuleHI{#7}\mdef\TabRuleHO{#7}\mdef\TabRuleHH{#7}%
		\mdef\TabRuleVI{#3}\mdef\TabRuleVO{#3}%
		\mdef\TabJustHO{#1}\mdef\TabJustHH{#1}%
		\SGMPTabJustHS{#4}%
		\SGMPTabRuleVS{#2}]%
	\let\SGMPnewline=\newline
}

\def\newline{\relax
	\ifvmode
		\vskip\baselineskip
	\else
		\unskip\vadjust{}\nobreak\hfil\break\vadjust{}\ignorespaces
	\fi
}

\def\SGMPTabJustHS#1{\SGMPTabcnvtlist{#1}\TabJustHS}

\def\SGMPTabRuleVS#1{\SGMPTabcnvtlist{#1}\TabRuleVS

\expandafter\SGMPrminitialdblsh\TabRuleVS\\\\\\\endSGMPrminitialdblsh\TabRuleVS
}

\def\SGMPrminitialdblsh\\#1\\\\#2\endSGMPrminitialdblsh#3{\def#3{#1}}

\documentstyle[12pt]{article}

\begin{document}

\title{Power law tail in the radial growth probability distribution for DLA}
\author{Peter Ossadnik and Jysoo Lee\\
H\"ochstleistungsrechenzentrum (HLRZ)\\
Forschungszentrum J\"ulich GmbH\\
Postfach 1913, W-5170 J\"ulich, Germany}
\maketitle

\begin{abstract}
\noindent Using both analytic and numerical methods, we study the radial growth
probability distribution $P(r,M)$ for large scale
off lattice diffusion limited aggregation (DLA) clusters.
If the form of $P(r,M)$ is a Gaussian, we show analytically that the width
$\xi(M)$ of the
distribution {\it can not} scale as the radius of gyration $R_G$ of the
cluster.
We generate about $1750$ clusters of masses $M$ up to $500,000$
particles, and calculate the distribution by sending $10^6$
further random walkers for each cluster.
We give strong support that the calculated distribution has a power law tail
in the interior ($r\sim 0$) of the cluster, and can be described by a scaling
Ansatz
$P(r,M) \propto {r^\alpha\over\xi}\cdot g\left( {r-r_0}\over \xi \right)$,
where $g(x)$ denotes some scaling
function which is centered around zero and has a width of order unity.
The exponent $\alpha$ is determined to be $\approx 2$, which is now
substantially smaller than values measured earlier.
We show, by including the power-law tail, that the width {\it can} scale as
$R_G$,
if $\alpha > D_f-1$.
\end{abstract}

\def\XRefId{}\section{\SGMPlab\XRefId Introduction}

\par The growth of DLA \SGMPcite{witten_sander}{}
clusters can be described by a set of growth proba\-bilities
\(
{\left\LeftPost{cub}p_{i}\right\RightPost{cub}}\)
. Each of them describes the probability that site
\(
i\)
 is touched by the next incoming particle. The
determination of the
\(
p_{i}\)
 and their distribution has gained much interest
\SGMPcite{meakin_stanley_coniglio_witten,halsey_meakin_procaccia,amitrano_coniglio_diliberto,schwarzer_lee_bunde_havlin_roman_stanley}{}.
Numerical calculations in this field were done by solving exactly the
Laplace equation on a given DLA cluster
\SGMPcite{mandelbrot_evertsz_physica_a,schwarzer_lee_havlin_stanley_meakin}{}.
Unfortunately these calculations are limited by computer resources to
cluster masses around 50,000 particles. An alternative quantity, which
is more accessible by large scale simulations, is the integrated radial
growth probability
\(
P{\left\LeftPost{par}r,M\right\RightPost{par}}\)
. It describes the probability that the next incoming
particle touches the cluster of mass
\(
M\)
 at a distance
\(
r\)
 from the seed. This quantity has been measured by
Plischke and Racz \SGMPcite{plischke_racz}{} who
studied the first two moments of the distribution and fitted a Gaussian
behavior
\def\XRefId{}
\begin{equation}\SGMPlab\XRefId\vcenter{\halign{\strut\hfil#\hfil&#\hfil\cr
$\displaystyle{P{\left\LeftPost{par}r,M\right\RightPost{par}}\propto \exp
{\left\LeftPost{par}-{{{\left\LeftPost{par}r-r_{0}\right\RightPost{par}}^{%
2}}\over{2\SGMPmathgrk{x}^{2}}}\right\RightPost{par}}
.}$\cr
}}\end{equation}
Using 4000 clusters of masses up to 2,500 particles they obtain a power
law behavior of the center
\(
r_{0}\)
 and the width
\(
\SGMPmathgrk{x}\)
 of the Gaussian
\def\XRefId{}
\begin{equation}\SGMPlab\XRefId\vcenter{\halign{\strut\hfil#\hfil&#\hfil\cr
$\displaystyle{r_{0}\propto R_{G}\propto M^{\SGMPmathgrk{n}}
,\hskip 0.265em \SGMPmathgrk{x}\propto M^{\SGMPmathgrk{n}'}}$\cr
}}\end{equation}
with exponents
\(
\SGMPmathgrk{n}=1/D_{f}\approx 0.585,\hskip 0.265em \SGMPmathgrk{n}'\approx
0.48\)
, where
\(
D_{f}\)
 denotes the fractal dimension of the clusters and
\(
R_{G}\)
 is the radius of gyration. Later, using larger scale
simulations, Meakin and Sander \SGMPcite{meakin_sander}{} showed that the
exponent
\(
\SGMPmathgrk{n}'\)
 approaches
\(
\SGMPmathgrk{n}\)
 with increasing cluster mass and hereby raised the
question about the real behavior of the width
\(
\SGMPmathgrk{x}\)
. Thus, currently three different possibilities are
being considered.
\SGMPbeginList{numeral}{}
\SGMPitem\def\XRefId{}\SGMPlab\XRefId  The width scales with the same exponent
\(
\SGMPmathgrk{n}\)
 as
\(
R_{G}\)
:
\(
\SGMPmathgrk{x}{\left\LeftPost{par}M\right\RightPost{par}}\propto
M^{\SGMPmathgrk{n}
}\)
{}.
\SGMPitem\def\XRefId{}\SGMPlab\XRefId The width scales with a smaller exponent
\(
\SGMPmathgrk{n}'\)
:
\(
\SGMPmathgrk{x}{\left\LeftPost{par}M\right\RightPost{par}}\propto
M^{\SGMPmathgrk{n}
'},\hskip 0.265em \SGMPmathgrk{n}'{\ifmmode<\else$<$\fi}\SGMPmathgrk{n}\)

\SGMPitem\def\XRefId{}\SGMPlab\XRefId The width scales with
\(
\SGMPmathgrk{n}\)
 but has logarithmic corrections:
\(
\SGMPmathgrk{x}{\left\LeftPost{par}M\right\RightPost{par}}\propto
M^{\SGMPmathgrk{n}
}/{\left\LeftPost{par}\ln {\left\LeftPost{par}M\right\RightPost{par}}
\right\RightPost{par}}^{\SGMPmathgrk{b}}\)
\SGMPendList
Especially case (3) has attracted some interest
\SGMPcite{coniglio_zannetti,ossadnik_physica_a_1991}{}. For an
exponent
\(
\SGMPmathgrk{b}=1/2\)
 it implies a multiscaling behavior of the mass
density
\(
M{\left\LeftPost{par}x\right\RightPost{par}}\propto r^{D{\left\LeftPost{par}
x\right\RightPost{par}}-1}\)
 where
\(
D{\left\LeftPost{par}x\right\RightPost{par}}\)
 is a non constant function of
\(
x=r/R_{G}\)
.\par

\par In the following we will show that case (1) combined with the
assumption of a Gaussian distribution leads to an unphysical singularity
in the mass distribution in the limit of infinite
\(
M\)
. This result is a consequence of the fact that the
Gaussian does not drop fast enough at the center of the cluster, which
suggests that
\(
\SGMPmathgrk{n}'=\SGMPmathgrk{n}\)
 cannot be true.\par

\par But on the other hand, we find numerically that a Gaussian is not a
good description for the growth probabilities near the seed of the
cluster, since for small
\(
r\)
 the tail behaves like a power law \SGMPcite{meakin_coniglio_stanley_witten}{}.
Furthermore one has
to realize that
\(
P{\left\LeftPost{par}r,M\right\RightPost{par}}\)
 drops to zero for large
\(
r\)
 simply because of the finite size of the cluster.
Usually such an effect would be taken into account by considering a
finite size cutoff function.\par

\par Here, we study the growth probabilities
\(
P{\left\LeftPost{par}r,M\right\RightPost{par}}\)
. We will show that the numerical data is only
consistent with a Gaussian behavior around the maximum but that this
description indeed fails in the small
\(
r\)
 tail. We will show that another type of behavior
\(
P{\left\LeftPost{par}r,M\right\RightPost{par}}\propto {{r^{\SGMPmathgrk{a}
}}\over{\SGMPmathgrk{x}}}{\ifmmode\cdot\else\.\fi}g{\left\LeftPost{par}{{%
r-r_{0}}\over{\SGMPmathgrk{x}}}\right\RightPost{par}}
\)
 where
\(
g{\left\LeftPost{par}x\right\RightPost{par}}\)
 is a scaling function is consistent with our
data as well around the maximum as in the tail. When this power law term
is included in the distribution the previously mentioned singularity
disappears.\par

\par Thus, the organization of the paper is as follows. In section 2 we
analyze the consequences of the different possible behaviors of
\(
\SGMPmathgrk{x}{\left\LeftPost{par}M\right\RightPost{par}}\)
 on the mass density. In section 3 we study
numerically the behavior of the growth probabilities and in section 4 we
show the consequences of the power law tail on the mass density.\par

\def\XRefId{Analytical_results}\section{\SGMPlab\XRefId Analytical results}

Following [7], we define two lengths, the radius $
r_0(M)$ and the width $\xi(M)$, as $ r_0(M) \equiv \langle r \rangle$,
 $\xi^2(M) \equiv \langle r^2 \rangle - \langle r \rangle ^2$, and $r = |\vec r
|$.
Here, $\langle f(r) \rangle \equiv \int f(r) P(r,M) dr$.
Let us assume that $P(r,M)$ can be described by a Gaussian distribution,
\begin{equation}
P(r,M) = {1 \over \sqrt {2 \pi \xi^2(M)}} \exp [-(r- r_0(M))^2 /
2\xi^2(M)].
\end{equation}
Here, we check whether the above form of $P(r,M)$ is consistent with
the growth of a fractal. We start from the relation
\begin{equation}
N(r,M) = \int_{1}^{M} P(r,M') dM',
\end{equation}
\noindent
where $N(r,M)dr$ is the number of sites within $[r,r+dr]$ in clusters of mass
$<M$.
Equation
(4) can be derived from the fact that any site that lies within $[r,r+dr]$,
in a cluster of mass $M$, has to be attached at the previous stage of
growth ($1 < M' < M$) [9].

Case (1): Let us assume that $ r_0(M) \equiv A M^{\nu}$ and $\xi(M) \equiv
BM^{\nu}$,
where $A, B$ are constants and we neglected other corrections. Using Eq.~(4),
we get
\begin{eqnarray}
N(r,M)
& = {1 \over \sqrt {2 \pi B^2}} \int_{1}^{M} {1 \over
M'^{\nu}} {\rm exp} [-(r-AM'^{\nu})^2 / (2B^2M'^{2 \nu})]dM' & \nonumber \\
& = {1 \over \sqrt {2 \pi B^2}} \exp (-{A^2 \over 2B^2}) \int_{1}^{M}
{1 \over M'^{\nu}} \exp (-{r^2 \over 2B^2M'^{2\nu}} + {rA \over B^2
M'^{\nu}})dM'. &
\end{eqnarray}
\noindent
Changing the variable of integration to $x \equiv M',^{-\nu}$ Eq.~(5)
becomes
\begin{equation}
N(r,M)
= {1 \over \nu \sqrt {2 \pi B^2}} \exp (-{A^2 \over 2B^2})
\int_{M^{-\nu}}^{1} {1 \over x^{1/\nu}} \exp (-{r^2x^2 \over 2B^2}
+ {rAx \over B^2})dx.
\end{equation}
\noindent
The total mass contained in the $[r,r+dr]$ shell is, by definition,
$N(r,M)dr$. For a fixed value of $r$, $N(r,M)$ becomes larger for
larger values of $M$. Especially, in the $M \to \infty$ limit, the
integral diverges due to the singularity at $x=0$ since $1/\nu$ is larger
than unity. This divergence
is inconsistent with the well-established fractal description of
DLA.
\bigskip

Case (2): Consider the case $\xi \equiv BM^{y \nu}$ with $0< y
<1$. The integral that corresponds to Eq.~(5) becomes
\begin{equation}
N(r,M)
= {1 \over \sqrt {2 \pi B^2}} \int_{1}^{M} {1 \over M'^{y \nu}}
\exp [-(r-AM'^{\nu})^2 / (2B^2M'^{2 y \nu})]dM'.
\end{equation}
\noindent
Changing the integration variable to $x \equiv M'^{-\nu}$, we get
\begin{equation}
N(r,M)
= {1 \over \nu \sqrt {2 \pi B^2}} \int_{M^{-\nu}}^{1} x^{y - 1 -
1/\nu}  \exp (-{A^2 \over 2B^2}x^{2(y-1)} - {r^2 \over 2B^2}x^{2y}
+ {rA \over B^2}x^{2y-1})dx.
\end{equation} \noindent
In the $M \to \infty$ limit, Eq.~(8) becomes
\begin{eqnarray}
N(r,M)
& = {1 \over \nu \sqrt {2 \pi B^2}} \int_{0}^{1} x^{y - 1
- 1/\nu}  {\rm exp} [-{x^{2y-2} \over 2B^2} (rx-A)^2]dx & \nonumber \\
& \propto r^{1 / \nu -1}. &
\end{eqnarray} \noindent
Since $D_f \equiv 1 / \nu$, Eq.~(9) becomes $N(r,M) \sim r^{D_f-1},$
consistent with the fact that the cluster is fractal with dimension
$D_f$.
\bigskip

Case (3): Finally, consider the case $\xi \equiv BM^{\nu}/ \sqrt{\ln
M}$.
\begin{equation}
N(r,M) = {1 \over \sqrt {2 \pi B^2}} \int_{1}^{M} {\sqrt{\ln M'} \over
M'^\nu} M'^{-{1 \over 2B^2} (rM'^{-\nu} -A)^2} dM'
\end{equation}
\noindent
Changing the variable to $x \equiv M'^{-\nu}$, Eq.~(10) becomes
\begin{equation}
N(r,M)
= {1 \over \sqrt {2 \pi \nu^3 B^2}} \int_{M^{-\nu}}^{1}
\sqrt {-\ln x}~x^{-1/\nu + (rx-A)^2/(2\nu B^2)} dx.
\end{equation}
\noindent
Since the logarithmic correction to the width appears not only in the
normalization of the Gaussian but also in its exponent, the behavior of the
integrand is now substantially changed as compared to case (1). There, one
observes
a behavior $x^{-1/\nu}\cdot \exp(-ax^2+bx+c)$ (Eq. (6)), whilst here one
obtains
$x^{-1/\nu+ax^2+bx+c}$.
The integral in Eq.~(11) can not be evaluated in a closed form, but we
can still extract some of the properties. First, consider the $M \to \infty$
limit. The logarithmic term in the integrand is divergent
at $x=0$ if the exponent of the power law term
$(A^2-2B^2) / (2 \nu B^2)$ is smaller than $-1$.
Therefore the whole integral is convergent only
if $A^2 > 2(1-\nu) B^2$. If the integral in Eq.~(11) is convergent,
the convergent value can be estimated by treating the  $\ln x$ term as
a constant. The resulting form is consistent with the idea of
multiscaling.

The analysis presented above shows that if $P(r,M)$ is Gaussian, the
width of $P(r,M)$, $\xi(M)$, can not scale as $R_G$. The divergence in
case (1) suggests that there is not enough ``screening'' at the inside
of the cluster. Therefore, $\xi(M)$ should increase slower than $R_G$ as
$M$ is increased.

In the next section, we numerically check whether the $P(r,M)$ can well be
described by a Gaussian.
%
%

\def\XRefId{}\section{\SGMPlab\XRefId Numerical simulations}

\par To measure the growth probabilities
\(
P{\left\LeftPost{par}x,M\right\RightPost{par}}\)
 numerically (here we scale
\(
x=r/R_{G}\)
) we generate clusters of masses 10,000, 20,000,
50,000, 100,000, 200,000 and 500,000 particles. For each mass we
generate at least 150 different clusters. Only for the 100,000 particle
clusters we grow 1000 samples to have very good statistics and to see
the details of the distribution. The growth probabilities are obtained
in a static measurement: First we grow each cluster to the full size and
afterwards we probe its surface using another
\(
10^{6}\)
 random walkers. Using a fast algorithm \SGMPcite{ossadnik_physica_a_1991}{},
the generation of one
100,000 particle cluster and the measurement of
\(
P{\left\LeftPost{par}x,M\right\RightPost{par}}\)
 can be done within 90 minutes on a Sun SPARC station
2. To obtain the
\(
P{\left\LeftPost{par}x,M\right\RightPost{par}}\)
 we bin it for
\(
x\in {\left\LeftPost{sqb}0,2.5\right\RightPost{sqb}}\)
 at 64 equidistant points and count the number of
particles touching the cluster at a radius between
\(
x\)
 and
\(
x+dx\)
. Fig. \SGMPref{P_x} shows the growth probabilities
for all cluster masses in linear and logarithmic scale. One has to
notice the power law tail of
\(
P{\left\LeftPost{par}x,M\right\RightPost{par}}\)
 at small
\(
x\)
, which is a property that cannot be found for pure
Gaussians. To smoothen out the noise we show in fig. \SGMPref{IP_x}
the integrated distributions
\(
IP{\left\LeftPost{par}x,M\right\RightPost{par}}=\int _{0}
^{x}P{\left\LeftPost{par}x',M\right\RightPost{par}}\hskip 0.167em
dx'\)
. For small cluster masses  up to 100,000
particles  one determines a systematic decrease of the width. For
the largest cluster masses no such a clear systematic behavior can be
seen. The curves for the 200,000 and 500,000 particle clusters lie on
top of each other. The width
\(
\SGMPmathgrk{x}\)
 can be studied quantitatively with two methods.\par

\par The first one determines the width by measuring the first and
second moments of the data
\(
\SGMPmathgrk{x}/R_{G}=\Rad{{\left\LeftPost{ang}x^{2}\right\RightPost{ang}}
-{\left\LeftPost{ang}x\right\RightPost{ang}}^{2}}\DoRad
\)
 and the second consists in fitting a Gaussian to the
data using a nonlinear fitting routine. Both methods give comparable
results and one obtains
\def\XRefId{}
\begin{equation}\SGMPlab\XRefId\vcenter{\halign{\strut\hfil#\hfil&#\hfil\cr
$\displaystyle{\SGMPmathgrk{x}_{\hbox{\rm Moments}}\propto
R_{G}{\ifmmode\cdot\else\.\fi}
M^{-0.044{\ifmmode\pm\else$\pm$\fi}0.003}}$\cr
$\displaystyle{\SGMPmathgrk{x}_{\hbox{\rm Gauss}}\propto
R_{G}{\ifmmode\cdot\else\.\fi}
M^{-0.050{\ifmmode\pm\else$\pm$\fi}0.003}}$\cr
}}\end{equation}
We also check whether our data is consistent with case (3) and we find
the behavior
\def\XRefId{}
\begin{equation}\SGMPlab\XRefId\vcenter{\halign{\strut\hfil#\hfil&#\hfil\cr
$\displaystyle{\SGMPmathgrk{x}_{\hbox{\rm Moments}}\propto {{R_{%
G}}\over{\Rad{\ln \hskip 0.167em {\left\LeftPost{par}
1.07{\ifmmode\cdot\else\.\fi}M\right\RightPost{par}}}\DoRad }}.}$\cr
}}\end{equation}
In fig. \SGMPref{width} we show the data of the width
\(
\SGMPmathgrk{x}_{\hbox{\rm Moments}}/R_{G}\)
 as a function of the cluster mass in semi logarithmic
scale. The solid lines in this plot are the results obtained by fitting
a power law and a
\(
1/\Rad{\ln M}\DoRad \)
 law to our data. For cluster masses up to 200,000
particles one seems to have good agreement with the assumption of a
logarithmic correction. Thus, our data seems to be inconsistent with
case (1)
\(
\SGMPmathgrk{n}'=\SGMPmathgrk{n}\)
. For clusters up to 200,000 particles our data are
consistent with cases (2) and (3).\par

\par Here, we want to study in more detail the form of the inner part of
\(
P{\left\LeftPost{par}x,M\right\RightPost{par}}\)
. Since fig. \SGMPref{P_x} shows a clear power law
behavior for small
\(
x\)
, we assume a simple scaling law
\def\XRefId{Power_law}
\begin{equation}\SGMPlab\XRefId\vcenter{\halign{\strut\hfil#\hfil&#\hfil\cr
$\displaystyle{P{\left\LeftPost{par}x,M\right\RightPost{par}}\propto {{x^{%
\SGMPmathgrk{a}}}\over{\SGMPmathgrk{x}/R_{G}}}{\ifmmode\cdot\else\.\fi}f
{\left\LeftPost{par}{{x-x_{0}}\over{\SGMPmathgrk{x}/R_{%
G}}}\right\RightPost{par}}}$\cr
}}\end{equation}
where
\(
f{\left\LeftPost{par}x\right\RightPost{par}}\)
 is some scaling function. This type of behavior is
used to stress the power law behavior of the small
\(
x\)
 tail.\par

\par If one uses for
\(
f{\left\LeftPost{par}x\right\RightPost{par}}\)
 too simple a function like a Fermi function
 which becomes constant for small
\(
x\)
, one does not describe the full mass dependence of
\(
P{\left\LeftPost{par}x,M\right\RightPost{par}}\)
 correctly: one measures only apparent large exponents
\(
\SGMPmathgrk{a}\)
, which increase logarithmically with
\(
M\)
 (fig. \SGMPref{alpha}).\par

\par To take this mass dependence correctly into account we
 perform a data collapse to the scaling
form (\SGMPref{Power_law}). Here we use for the width
\(
\SGMPmathgrk{x}/R_{G}\)
 and center
\(
x_{0}\)
 the values calculated from the moments of
\(
P{\left\LeftPost{par}x,M\right\RightPost{par}}\)
 and vary the exponent $\alpha$. The best collapse of our
data, which is shown in fig. (\SGMPref{scaling_plot}), we obtain for a
value
\(
\SGMPmathgrk{a}=2.0{\ifmmode\pm\else$\pm$\fi}0.4\)
. This value of $\alpha$ is much smaller than those
obtained before and in \SGMPcite{meakin_stanley_coniglio_witten}{}. Thus, the
previously
measured large exponents are only {``}apparent{''} values. They
are combinations of the power law tail and the mass dependence of the
width of the scaling function. If one takes into account the mass
dependence of the scaling function the $\alpha$ decrease very much. Anyhow,
we still find a strong power law screening of the inner parts of the
cluster.\par

\def\XRefId{}\section{\SGMPlab\XRefId Consequences}

\par Having established the existence of the power-law tail of
\(
P{\left\LeftPost{par}r,M\right\RightPost{par}}\)
, we study how this modifies the results in Sect.
\SGMPref{Analytical_results}. Here we assume that
\(
P{\left\LeftPost{par}r,M\right\RightPost{par}}\)
 can be described by eq. (\SGMPref{Power_law}) with
a normalization factor
\(
N_{0}\)
. Let the average of
\(
f{\left\LeftPost{par}x\right\RightPost{par}}\)
 be zero and the width be some constant. We first
consider the case (1). Since
\(
P{\left\LeftPost{par}r,M\right\RightPost{par}}\)
 has to be normalized, it is
\def\XRefId{Int_1}
\begin{equation}\SGMPlab\XRefId\vcenter{\halign{\strut\hfil#\hfil&#\hfil\cr
$\displaystyle{\int _{0}^{\infty }{{N_{%
0}}\over{\SGMPmathgrk{x}}}{\left\LeftPost{par}{{r}
\over{R_{G}}}\right\RightPost{par}}^{\SGMPmathgrk{a}
}f{\left\LeftPost{par}{{r-r_{0}}\over{\SGMPmathgrk{x}
}}\right\RightPost{par}}\hskip 0.167em dr=1}$\cr
}}\end{equation}
Changing the integration variable to
\(
x=r/\SGMPmathgrk{x}\)
 eq. (\SGMPref{Int_1}) becomes
\def\XRefId{}
\begin{equation}\SGMPlab\XRefId\vcenter{\halign{\strut\hfil#\hfil&#\hfil\cr
$\displaystyle{N_{0}{\left\LeftPost{par}{{\SGMPmathgrk{x}}\over{R_{%
G}}}\right\RightPost{par}}^{\SGMPmathgrk{a}}{\ifmmode\cdot\else\.\fi}
\int _{0}^{\infty }x^{\SGMPmathgrk{a}}
f{\left\LeftPost{par}x-x_{0}\right\RightPost{par}}\hskip 0.167em
dx=1,}$\cr
}}\end{equation}
where
\(
x_{0}\equiv r_{0}/\SGMPmathgrk{x}.\)
 Since the integral is a constant,
\(
N_{0}\)
 should scale as
\(
{\left\LeftPost{par}R_{G}/\SGMPmathgrk{x}\right\RightPost{par}}^{%
\SGMPmathgrk{a}}\).
The equation that corresponds to (5) is
\def\XRefId{Int_2}
\begin{equation}\SGMPlab\XRefId\vcenter{\halign{\strut\hfil#\hfil&#\hfil\cr
$\displaystyle{N{\left\LeftPost{par}r,M\right\RightPost{par}}}$\hfilneg&$\displaystyle{{}\propto
\int _{1}^{M}{{1}\over{\SGMPmathgrk{x}
}}{\left\LeftPost{par}{{r}\over{\SGMPmathgrk{x}}}\right\RightPost{par}}^{%
\SGMPmathgrk{a}}f{\left\LeftPost{par}{{r-r_{0}}\over{%
\SGMPmathgrk{x}}}\right\RightPost{par}}\hskip 0.167em dM}$\cr
$\displaystyle{}$\hfilneg&$\displaystyle{{}\propto \int _{1}^{M}
M^{-{\left\LeftPost{par}1+\SGMPmathgrk{a}\right\RightPost{par}}/D_{f}
}r^{\SGMPmathgrk{a}}f{\left\LeftPost{par}{{r-r_{0}
}\over{\SGMPmathgrk{x}}}\right\RightPost{par}}\hskip 0.167em
dM}$\cr
}}\end{equation}
The part that causes a divergence in (5) is the
\(
M^{-{\left\LeftPost{par}1+\SGMPmathgrk{a}\right\RightPost{par}}/D_{f}
}\)
 term, since other terms in the integrand become
constants for a fixed value of
\(
r\)
. Therefore, the integral (\SGMPref{Int_2}) diverges
if the exponent
\(
{\left\LeftPost{par}1+\SGMPmathgrk{a}\right\RightPost{par}}/D_{f}\)
 is less than unity, which is equivalent to
\(
\SGMPmathgrk{a}{\ifmmode<\else$<$\fi}D_{f}-1\)
. Since the numerical value of
\(
\SGMPmathgrk{a}\)
 is obviously larger than
\(
\MthAcnt {\SGMPmathgrk{a}}{\tilde }=D_{f}-1\simeq 0.
71\)
, our Ansatz is not anymore excluded by the
self-consistency argument. One can show, in a similar way, that the
existence of a power-law tail is consistent with the other cases (2) and
(3).\par

\def\XRefId{}\section{\SGMPlab\XRefId Conclusion}

\par We have studied the growth probabilities for large off lattice DLA
clusters. We have shown that the usual assumption of a Gaussian behavior
leads to unphysical singularities under the assumption that the width of
the Gaussian scales with the same exponent as its center. We have shown
that extensive numerical calculations seem to be inconsistent with the
assumption
\(
\SGMPmathgrk{n}'=\SGMPmathgrk{n}\)
. It seems that cases (2) or (3) have to be favored.
Moreover, we showed that
\(
P{\left\LeftPost{par}r,M\right\RightPost{par}}\)
 has a power law tail which is inconsistent with the
Gaussian behavior that is usually assumed. Therefore we suggested
another type of behavior
\(
P{\left\LeftPost{par}r,M\right\RightPost{par}}\propto {{r^{\SGMPmathgrk{a}
}}\over{\SGMPmathgrk{x}}}f{\left\LeftPost{par}{{r-r_{%
0}}\over{\SGMPmathgrk{x}}}\right\RightPost{par}}\)
 which removes the unphysical singularity described
above. This type of behavior describes the maximum of the measured
growth probabilities as well as the tail. We showed that one explicitly
has to take into account the dependence of the width of the distribution
on the cluster mass in order to find the correct exponent
\(
\SGMPmathgrk{a}\approx 2.\)
 If one uses a simple ad hoc assumption do determine
the inner tail of the distribution one mixes the {``}real{''}
power law term and contributions from the scaling function.\par

\par Unfortunately our data does not enable us to draw any conclusions
about the behavior of the minimum growth probabilities, as studied for
example in \SGMPcite{schwarzer_lee_bunde_havlin_roman_stanley}{}. The data
we measured is only an average of the growth probabilities within one
{``}shell{''} of the cluster. Thus, we only have data about the
spatial distribution of growth probabilities and not about the
distribution
\(
n{\left\LeftPost{par}P\right\RightPost{par}}\)
 of the growth probabilities themselves. A measurement
of
\(
n{\left\LeftPost{par}P\right\RightPost{par}}\)
 is for small
\(
P\)
 probably not feasible with a random walker
method.\par

\par We gratefully acknowledge stimulating discussions with H. J.
Herrmann and D. Stauffer.\par

\newpage
\appendix\def\XRefId{}

\begin{figure}[h]
\caption{\def\XRefId{P_x}\SGMPlab\XRefId Growth probability distributions for
different cluster masses in linear and double logarithmic scale.}
\end{figure}

\begin{figure}[h]
\caption{\def\XRefId{IP_x}\SGMPlab\XRefId Integrated probability
distributions.}
\end{figure}

\begin{figure}[h]
\caption{\def\XRefId{width}\SGMPlab\XRefId Scaling of the width of the
distribution as obtained by measuring the moments of the distribution. The two
 solid lines denote the fits of a power law and a
$ 1/{\ln M} $ law to our data.}
\end{figure}

\begin{figure}[h]
\caption{\def\XRefId{alpha}\SGMPlab\XRefId Behavior of the {``}apparent{''}
exponents of the power law with varying cluster mass. This demonstrates,
 that the mass dependence is important in order to obtain the correct
exponent.}
\end{figure}

\begin{figure}[h]
\caption{\def\XRefId{scaling_plot}\SGMPlab\XRefId Plot of the scaling function
$g(y)\propto P(\xi\cdot y+x_0,M)\cdot\xi\cdot(\xi\cdot y+x_o)^\alpha$.
 The maximum of the distribution collapses as well as the power law tail.}
\end{figure}

\end{document}